\documentclass[a4paper,conference]{IEEEtran}

\usepackage{cite}
\usepackage{graphicx}

\usepackage{amsmath,amssymb}   
\usepackage{aas_macros,bm}

\usepackage{fancyheadings}
\begin{document}

\title{MULTIGRAIN: Simulating mixtures of multiple dust grains and gas with SPH}

\author{\IEEEauthorblockN{Daniel J. Price}
\IEEEauthorblockA{School of Physics \& Astronomy \\
Monash University\\
Vic. 3800, Australia\\
daniel.price@monash.edu}
\and
\IEEEauthorblockN{Mark A. Hutchison}
\IEEEauthorblockA{Physikalisches Institut\\
Universit\"at Bern\\
Gesellschaftstrasse 6\\
3012 Bern, Switzerland
}
\and
\IEEEauthorblockN{Guillaume Laibe}
\IEEEauthorblockA{University of Lyon, Univ Lyon1, Ens de Lyon, \\
CNRS, Centre de Recherche Astrophysique de Lyon \\
UMR5574, F-69230, Saint-Genis-Laval \\
France}}

\maketitle

\begin{abstract}
We present MULTIGRAIN, an algorithm for simulating multiple phases of small dust grains embedded in a gas, building on our earlier work in simulating two-phase mixtures of gas and dust in SPH \cite{laibeprice12,laibeprice12a,pricelaibe15}. The MULTIGRAIN method \cite{hpl18} is more accurate than single-phase simulations because the gas experiences a backreaction from each dust phase and communicates this change to the other phases, thereby indirectly coupling the dust phases together. The MULTIGRAIN method is fast, explicit and low storage, requiring only an array of dust fractions and their derivatives defined for each resolution element. We demonstrate the MULTIGRAIN algorithm on test problems related to the settling of dust in the discs of gas around young stars, where solar systems are born.  Finally I will discuss possible extensions of the method to incorporate both large and small grains, together with recent improvements in our numerical techniques for gas and dust mixtures. In particular, I will show how the `overdamping' problem identified by \cite{laibeprice12} can be solved.
\end{abstract}

\section{Introduction}
 Dust you are, and to dust you shall return. You, the ground you stand on, and the Earth itself, are all made of cosmic dust. How did this material assemble in interstellar space in order to form planets like the Earth? 
 
 Since Laplace's nebular hypothesis of 1755 we have understood that planets are born in swirling discs of gas and dust surrounding newborn stars. Interstellar dust is the belly-button fluff of the universe: the detritus of stars. Much like the household version under your bed, it consists of the wind-blown products of stellar living --- silicate and carbon-rich material that are the basic building blocks of planets.
 
 We are interested in modelling the formation of planets in the discs of gas orbiting around newborn stars. Much like the different grains in a loaf of bread, dust in protoplanetary discs is a mix of grains of difference sizes. The main numerical issue is that these grains respond to the gas in very different ways. Large grains --- anything up to km-sized planetesimals --- move in a way that is completely decoupled from the gas. Small grains --- from sub-micron to micron sizes --- largely stick to the gas. Grain sizes in between are in the `sweet spot' where they rapidly settle and drift in radius towards the star \cite{weidenschilling77}. 
 
 Settling towards the disc midplane occurs because dust grains, unlike gas molecules, do not collide sufficiently often to generate pressure. Radial drift occurs because solid particles aim to orbit the star at the true Keplerian speed, whereas gas orbits slower because of the radial pressure gradient. This means that grains feel a headwind as they orbit through the gas, producing a drift towards the star. This presents a problem for planet formation since grains in the sweet spot are rapidly hoovered into the star, leaving no raw material left with which to form planets.
  
  A further complication is that the dust motion --- by conservation of momentum --- has a backreaction on the gas. If the dust migrates towards the star, the gas must migrate away from the star. The effect is usually thought to be small because the interstellar dust mass is typically 1\% of the gas mass, but dust fractions can be significantly higher than this in protoplanetary discs (even of order unity in the disc midplane). Backreaction of the dust on the gas may assist planet formation, for example by enabling self-induced dust traps \cite{glm17} and streaming instabilities \cite{youdingoodman05}. This means it is not sufficient to model a single phase of dust interacting with the gas, because \emph{the gas responds to the bulk motion of the entire grain population}. This motivates our new algorithm.
  
  Here we present the most recent update of our attempts to model dust-gas mixtures in Smoothed Particle Hydrodynamics (SPH), by extending our previous single-phase dust-gas SPH algorithm to a population of dust grains of different sizes. For the first time we are able to study the backreaction on the gas from a population of small grains self-consistently. The paper is structured as follows: In Section~\ref{sec:single} we discuss the main issues related to simulating single-phase dust-gas mixtures. Section~\ref{sec:sphsingle} presents our current approach to single grain simulations in SPH. In Section~\ref{sec:multi} we show how to generalise the SPH algorithm to the case where there are multiple dust grains. Finally, we present numerical tests of our new {\rm multigrain} algorithm.
    
\section{Physics of single phase dust-gas mixtures}
\label{sec:single}
\subsection{Equations for single phase mixture}
 The basic equations for gas-dust mixtures are easy to write down, just the equations of hydrodynamics for two components (dust and gas) expressing the conservation of mass, momentum and energy:
\begin{align}
\frac{\partial \rho_{\rm g}}{\partial t} + (\bm{v}_{\rm g}\cdot \nabla) \rho_{\rm g}  & =  - \rho_{\rm g} (\nabla \cdot \bm{v}_{\rm g}), \label{eq:rhog} \\
\frac{\partial \rho_{\rm d}}{\partial t} + (\bm{v}_{\rm d}\cdot \nabla) \rho_{\rm d}  & =  - \rho_{\rm d} (\nabla \cdot \bm{v}_{\rm d}), \\
\frac{\partial \bm{v}_{\rm g}}{\partial t} + (\bm{v}_{\rm g}\cdot \nabla) \bm{v}_{\rm g}  & = - \frac{\nabla P_{\rm g}}{\rho_g} - \nabla \Phi + \frac{K}{\rho_{\rm g}}( \bm{v}_{\rm d} -  \bm{v}_{\rm g}), \label{eq:vg} \\
\frac{\partial \bm{v}_{\rm d}}{\partial t} + (\bm{v}_{\rm d}\cdot \nabla) \bm{v}_{\rm d}  & =  - \nabla \Phi - \frac{K}{\rho_{\rm d}}( \bm{v}_{\rm d} -  \bm{v}_{\rm g}), \label{eq:vd} \\
\frac{\partial u_{\rm g}}{\partial t} + (\bm{v}_{\rm g}\cdot \nabla) u_{\rm g}  & =  - \frac{P_{\rm g}}{\rho_{\rm g}}(\nabla \cdot \bm{v}_{\rm g}) + \frac{K}{\rho_{\rm g}}( \bm{v}_{\rm d} -  \bm{v}_{\rm g})^2, \label{eq:ug}
\end{align}
where $\rho_{\rm g}, \rho_{\rm d}, \bm{v}_{\rm g}$ and $\bm{v}_{\rm d}$ are the gas and dust densities and velocities, respectively, $u_{\rm g}$ is the specific thermal energy of the gas and $\Phi$ is the gravitational potential. Dust and gas exchange momentum via aerodynamic drag, the physics of which is encapsulated in the drag coefficient $K$. Typically $K$ is inversely proportional to grain size, so that strong drag occurs for small grains and weak drag occurs on large grains.

\subsection{The stopping time}
 Drag introduces a new timescale. Consider a mixture with uniform densities and velocities. Subtracting equation (\ref{eq:vg}) from equation (\ref{eq:vd}) gives
\begin{equation}
\frac{\partial (\bm{v}_{\rm d} - \bm{v}_{\rm g})}{\partial t} =  -\frac{( \bm{v}_{\rm d} -  \bm{v}_{\rm g})}{t_{\rm s}}, \label{eq:dv}
\end{equation}
where
\begin{equation}
t_{\rm s} \equiv \frac{\rho_{\rm d}\rho_{\rm g}}{K (\rho_{\rm d} + \rho_{\rm g})}
\end{equation}
is the \emph{stopping time}. From (\ref{eq:dv}) we see that $t_{\rm s}$ is the characteristic timescale on which the differential velocity of the mixture decays. That is, the timescale for the two phases of the mixture to glue each other together.

 For simulations involving dust, the relevant dimensionless parameter is the ratio of the stopping time to the typical timescale of interest. In protoplanetary discs the relevant timescale is the orbital period $t_{\rm orb} = 2\pi/\Omega$, where $\Omega = \sqrt{GM/R^3}$ is the Keplerian angular velocity for an orbit at radius $R$ around a star of mass $M$. The resultant dimensionless parameter $S_{\rm t} \equiv t_s \Omega$ is known as the \emph{Stokes number}. Therefore, $S_{\rm t} = 1$ is the sweet spot, $S_{\rm t} \ll 1$ grains remain stuck to the gas, and $S_{\rm t} \gg 1$ grains are effectively decoupled. For the rest of the proceedings grains with Stokes numbers less than or greater than unity are what we refer to as `small' or `large' grains, respectively. 
 

\subsection{To infinity and beyond}
 The difficult aspect for numerical codes is that the stopping time can range from zero to infinity, and these are both perfectly sensible and well-behaved limits. This is best illustrated by considering  a linear perturbation analysis of equations (\ref{eq:rhog})--(\ref{eq:ug}), which leads to the dispersion relation \cite{priceetal18a}
\begin{equation}
(\omega^2 - c_{\rm s}^2 k^2) + \frac{i}{t_{\rm s}\omega} (\omega^2 - \tilde{c}_{\rm s}^2 k^2) = 0. \label{eq:disp}
\end{equation}
The limit $t_{\rm s}\to \infty$ corresponds to undamped sound waves in the gas ($\omega = \pm k c_{\rm s}$) propagating at the sound speed, $c_{\rm s}$. The limit $t_{\rm s} \to 0$ produces undamped sound waves in the \emph{mixture} ($\omega = \pm k \tilde{c}_{\rm s}$), propagating at the \emph{modified sound speed}, $\tilde{c}_{\rm s} \equiv c_{\rm s} \sqrt{1 + \rho_{\rm d}/\rho_{\rm g}}$, representing a gas with extra inertia from the carried dust. Waves are most strongly damped when the stopping time is comparable to the wave period, $t_{\rm s} \sim 1/\omega$.

\section{Single phase dust-gas mixtures in SPH}
\label{sec:sphsingle}

\subsection{Large grains: Use two sets of particles}
 The simplest method of discretising (\ref{eq:rhog})--(\ref{eq:ug}) in SPH is to represent the mixture with two sets of SPH particles, one set for gas and one for dust \cite{monaghankocharyan95}. The discrete equations are then simply the usual SPH equations implemented for each phase, except that there is no acceleration on dust particles from the pressure gradient. Gas particle quantities are summed over gas neighbours, and dust particle quantities are summed over dust neighbours, straightforwardly generalising the usual SPH density sum \cite{laibeprice12}. 
 
 Our implementation of this method was described in \cite{laibeprice12,laibeprice12a} and was presented in the 2015 SPHERIC proceedings. Only the drag terms require interpolation over the particles of the \emph{other} type. We discretise these using
 \begin{align}
\left. \frac{{\rm d} \bm{v}^a_{\rm g}}{{\rm d} t} \right\vert_{\rm drag} =
& + \nu \sum_i \frac{m_i}{\rho_i \rho_a t^{aj}_{\rm s}} \left[\bm{v}_{ai}  \cdot \hat{\bm{r}}_{ai} \right]\hat{\bm{r}}_{ai} D_{ai} (h_a), \label{eq:vgsph} \\
\left. \frac{{\rm d} \bm{v}^i_{\rm d}}{{\rm d} t} \right\vert_{\rm drag} = & - \nu \sum_a \frac{m_a}{\rho_a \rho_i t^{ai}_{\rm s}}  \left[\bm{v}_{ai}  \cdot \hat{\bm{r}}_{ai} \right]  \hat{\bm{r}}_{ai} D_{ai} (h_a), \label{eq:vdsph}\\
\left. \frac{{\rm d} u^a_{\rm g}}{{\rm d} t} \right\vert_{\rm drag} = & \nu \sum_i \frac{m_i}{\rho_a \rho_i t^{ai}_{\rm s}}  \left[\bm{v}_{ai}  \cdot \hat{\bm{r}}_{ai}  \right]^2  D_{ai} (h_a) \label{eq:dusph}
\end{align}
where $a$ and $b$ refer to gas particles, $i$ and $j$ refer to dust particles, $\nu$ is the number of dimensions, and $D_{ai}$ is a \emph{double hump kernel}, which we found to produce an order of magnitude better accuracy when solving (\ref{eq:dv}) compared to using a Bell-shaped kernel, at no extra cost \cite{laibeprice12}. The discrete stopping time is defined via
\begin{equation}
t_{\rm s}^{aj} \equiv \frac{\rho^a_{\rm g} \rho^j_{\rm d}}{K (\rho^a_{\rm g} + \rho^j_{\rm d})}.
\end{equation}
For the purpose of testing one may assume $K$ is constant but in our 3D code $t_{\rm s}$ is set according to a physical drag law. 

\subsection{Problems with using two sets of particles at high drag}
\label{sec:problems}
 In our original investigation \cite{laibeprice12} we found that while discretising the dust-gas equations as above was straightforward when the drag is weak (i.e., for large grains), the method could be woefully inaccurate when the drag is strong (i.e., when simulating small grains). Elementary considerations already suggest that one cannot simply employ (\ref{eq:vgsph})--(\ref{eq:dusph}) in the limit where $K\to \infty$ ($t_{\rm s} \to 0$), since the denominators go to zero and the terms themselves become infinite. We found not only the usual requirement for the computational timestep to be shorter than the stopping time ($\Delta t < t_{\rm s}$; which can be remedied with implicit methods \cite{laibeprice12a}; see also Monaghan this proceedings), but also a \emph{spatial resolution requirement} to prevent artificial overdamping of waves, namely
\begin{equation}
h \lesssim c_{\rm s} t_{\rm s}, \label{eq:hcsts}
\end{equation}
implying that an infinite number of SPH particles are required in the limit where $t_{\rm s} \to 0$.

The resolution problem was borne out in our numerical tests. For example, we found that in wave and shock problems with $t_{\rm s} \sim 1/500$ of the wave period we required as many as 10,000 particles in 1D to match the analytic solution given by (\ref{eq:disp}) \cite{laibeprice12}. We dubbed this the `overdamping problem'. We have recently found a solution for it which we present in Section~\ref{sec:overdamping}. 

Problems also arise from different resolution lengths. For example, we found that dust particles could artificially concentrate and become trapped below the resolution of the gas \cite{laibeprice12a}, leading to potentially misleading conclusions about the degree of dust clumping seen in simulations of interstellar turbulence (see \cite{tpl17}). Our best approach to this at present is over-resolve the gas compared to the dust. That is, we employ many more gas particles than dust particles in our simulations (e.g. \cite{priceetal18}), though this is not a fail-safe procedure.
 

\subsection{What God has joined together, let man not separate}
 In \cite{laibeprice14} we showed that all of the above problems could be solved --- in the limit where the grains are small --- by a change of variables and perspective. Rather than consider gas and dust as two separate phases, in the small grain limit it makes better sense to view  \emph{the mixture as a whole} and consider only the differential forces which separate the two phases.
 
  The change of variables is straightforward. We define a new set of variables 
\begin{align}
\rho & \equiv \rho_{\rm g} + \rho_{\rm d}, \\
\epsilon & \equiv \rho_{\rm d}/\rho, \\
\bm{v} & \equiv (\rho_{\rm g} \bm{v}_{\rm g} + \rho_{\rm d} \bm{v}_{\rm d})/(\rho_{\rm g} + \rho_{\rm d}), \\
\Delta \bm{v} & \equiv \bm{v}_{\rm d} - \bm{v}_{\rm g} \\
\tilde{u} & \equiv (1 - \epsilon) u_{\rm g},
\end{align}
representing the total density, dust fraction, barycentric velocity, differential velocity and thermal energy, respectively. Equations (\ref{eq:rhog})--(\ref{eq:ug}) can then be rewritten in terms of this new set of variables \cite{laibeprice14}.

 The change in perspective is that we now solve the equations using \emph{only one set of SPH particles} which move with the barycentric velocity, $\bm{v}$. Rather than representing gas or dust, these particles now represent the mixture. This means that the dust fraction $\epsilon$ and the differential velocity $\Delta \bm{v}$ are now \emph{intrinsic} properties, rather than representing actual differences in density or velocity between different sets of particles.

 Deriving the equations in our new set of variables can be done without approximation and in \cite{laibeprice14a} we derived an SPH algorithm based on discretising these `full one-fluid' equations. We found that this indeed solved the overdamping and mixed resolution problems. The caveat is that by using a single set of particles we implicitly assume that the dust is a fluid, excluding effects that are important when grains are large --- namely that dust particles should be free to stream through each other in opposite directions. In the one fluid method, the mixture particles are not allowed to penetrate each other so the method becomes inaccurate in this regime. But this is precisely where the two-sets-of-particles method is simple, accurate and explicit! So the mixture formulation should only be applied when the drag is strong (i.e., for small grains).
 
\subsection{Small grains: Use one set of mixture particles}
  It turns out that the mixture equations simplify greatly if one assumes that the stopping time is short compared to the timescales of interest. In this limit, known as the \emph{terminal velocity approximation}, the differential velocity is prescribed according to
\begin{equation}
\Delta \bm{v} \approx t_{\rm s} \frac{\nabla P}{\rho_{\rm g}},
\end{equation}
and the equations for the mixture become
\begin{align}
\frac{{\rm d}\rho}{{\rm d}t} & = -\rho (\nabla\cdot\bm{v}),  \label{eq:drhodt} \\
\frac{{\rm d}\bm{v}}{{\rm d}t} & = -\frac{\nabla P}{\rho} - \nabla \Phi, \\
\frac{{\rm d}\tilde{u}}{{\rm d}t} & = -\frac{P}{\rho} (\nabla\cdot\bm{v}).  \label{eq:dudt} \\
\frac{{\rm d}\epsilon}{{\rm d}t} & = -\frac{1}{\rho} \nabla \cdot (\epsilon t_{\rm s} \nabla P), \label{eq:depsdt}
\end{align}
These are just the usual fluid equations supplemented by an evolution equation for the dust fraction! Equation (\ref{eq:depsdt}) reveals the physics of dust in a beautifully intuitive way --- pressure gradients drive changes in the dust-to-gas ratio, which is precisely how we think about settling and radial drift in protoplanetary discs (considering pressure gradients either in $z$ or $R$, respectively). The limit where $t_{\rm s} \to 0$ is also now trivial --- this simply reduces to the usual equations of fluid dynamics with constant $\epsilon$. The modified sound speed $\tilde{c}_{\rm s}$ that should appear --- by considering (\ref{eq:disp}) when $t_{\rm s} \to 0$ --- arises because the pressure depends only on the \emph{gas} density, hence the equation of state for an adiabatic gas is given by
\begin{equation}
P = (\gamma - 1) \rho_{\rm g} u_{\rm g} = (\gamma - 1) (1 - \epsilon) \rho u_{\rm g}.
\end{equation}

In \cite{pricelaibe15} we presented an algorithm to solve (\ref{eq:drhodt})--(\ref{eq:depsdt}) in SPH. Here we generalise this algorithm to the case where there are multiple populations of small grains.

\section{Multiphase dust-gas mixtures in SPH}
\label{sec:multi}

 Generalising the above method to account for multiple populations of dust grains is relatively straightforward. In \cite{laibeprice14b} we derived the equations for a multiple-dust-grains-and-gas mixture without approximation (that is, without assuming that the stopping time is small). For the SPH implementation we presently consider the case where the stopping time is short. Description of our \textsc{multigrain} algorithm and corresponding tests can be found in \cite{hpl18}.

\subsection{Mixture equations and definitions}
Equations (\ref{eq:drhodt})--(\ref{eq:depsdt}) generalised to the case where there are multiple grain populations $j=1,2,3...N$ are given by
\begin{align}
\frac{{\rm d} \rho}{{\rm d} t}& =  - \rho \left( \nabla \cdot \bm{v} \right), \label{eq:drhodtmulti} \\
\frac{{\rm d} \bm{v}}{{\rm d} t} & = -\frac{\nabla P}{\rho} -\nabla \Phi, \label{eq:dvdtmulti} \\
\frac{{\rm d} \tilde{u}}{{\rm d} t} & = - \frac{P}{\rho} (\nabla \cdot \bm{v}), \label{eq:dudtmulti} \\
\frac{{\rm d} \epsilon_j}{{\rm d} t} & =  - \frac{1}{\rho} \nabla \cdot ( \epsilon_j \tilde{t}_{{\rm s}, j} \nabla P ). \label{eq:depsjdt}
\end{align}
Hidden in the above are a swag of definitions designed to make the equations appear similar to the single phase equations. For example, we define $\tilde{u} \equiv (1 - \epsilon) u_{\rm g}$ as previously, but $\epsilon$ is now the \emph{total} dust fraction, given by the sum of the fractions of the individual phases
\begin{equation}
\epsilon \equiv \frac{\rho_{\rm d}}{\rho} = \sum_j \epsilon_j.
\end{equation}
Likewise, the density is now the sum of the gas density and the density of all the individual dust phases
\begin{equation}
\rho \equiv \rho_{\rm g} + \sum_j \rho_{{\rm d}, j}
\end{equation}
and similarly for the barycentric velocity
\begin{equation}
\bm{v} \equiv \left( \rho_{\rm g} \bm{v}_{\rm g} + \sum_j \rho_{{\rm d}, j} \bm{v}_{{\rm d}, j} \right)/\rho.
\end{equation}

We define the effective stopping time, $\tilde{t}_{{\rm s}, j}$, below.

\begin{figure*}
\begin{center}
\includegraphics[width=\textwidth]{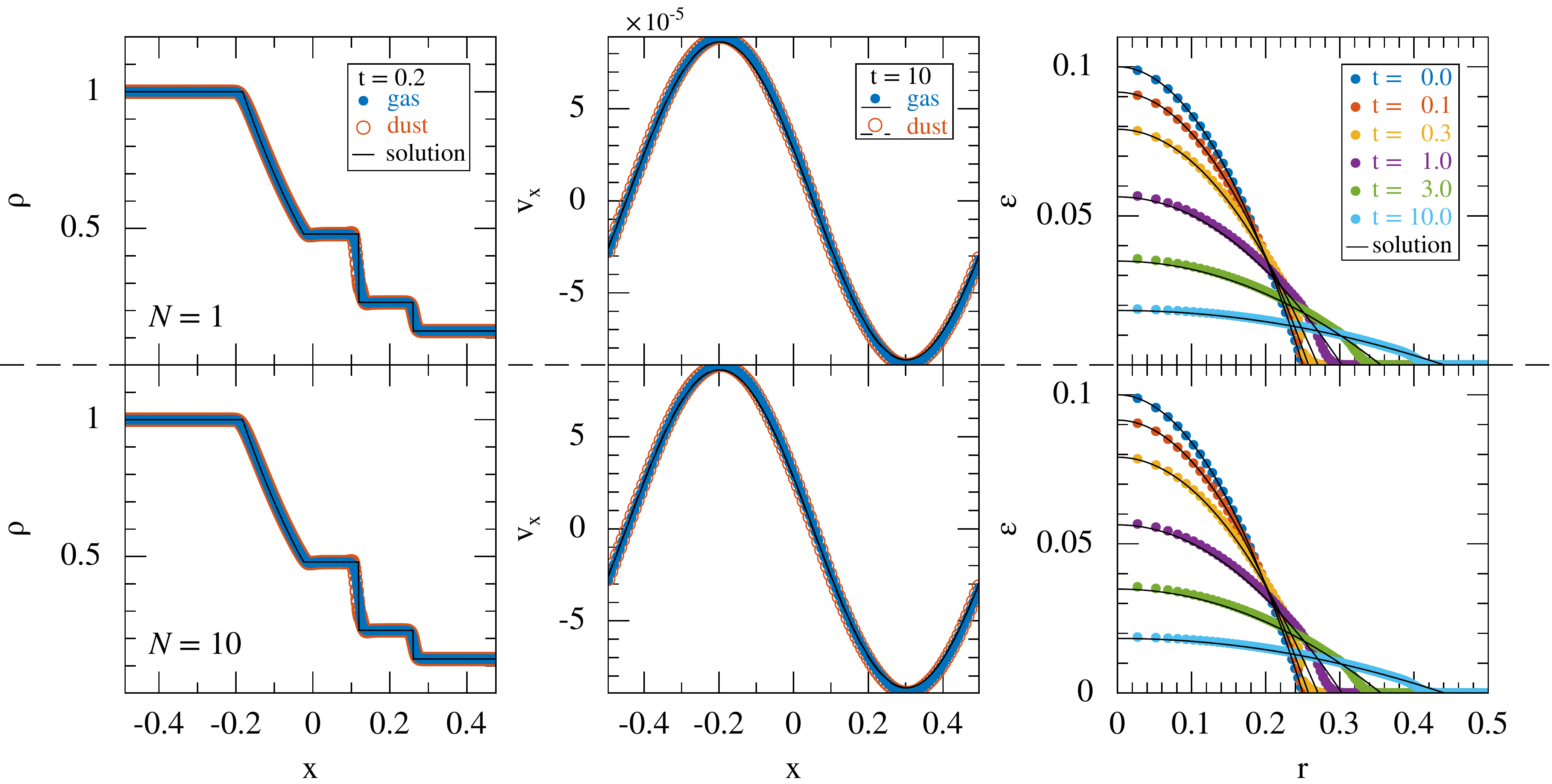}
\caption{Results of the {\sc dustyshock}, {\sc dustywave} and dust diffusion problems (left to right, respectively), comparing the results for a single grain size (top) to the results of a {\sc multigrain} simulation where we evolve 10 sub-populations of grains of the same size (bottom). This confirms that our algorithm self-consistently reduces to the single phase solutions when the grain sizes are equal, and that these match the corresponding analytic solutions (solid lines). Reproduced from \cite{hpl18}.}
\label{fig:dusttest}
\end{center}
\end{figure*}

\subsection{Stopping time when there are multiple dust species}
The most confusing definition relates to the stopping time, since there are now both individual `stopping times' corresponding to the interaction between each different dust phase and the gas and `effective stopping times' for the mixture as a whole. If one considers the original set of equations (\ref{eq:rhog})--(\ref{eq:ug}) with drag interaction terms for each species with a different drag coefficient $K_j$, then one may define a timescale for each dust species according to
\begin{equation}
t_{j} \equiv \frac{\rho}{K_j}.
\end{equation}
The effective stopping time required in the ${\rm d}\epsilon_j/{\rm d}t$ equation for each phase then turns out to be given by
\begin{equation}
\tilde{t}_{{\rm s}, j} \equiv \epsilon_j t_j - \sum_k \epsilon_k^2 t_{\rm k}. \label{eq:tsj}
\end{equation}
That is, one must consider the difference between the stopping time for each phase and the weighted sum of the stopping times of the other phases. 

One may also define an effective stopping time for the mixture in terms of the physical grain sizes. Here one first defines the effective grain size $s$ according to
\begin{equation}
s \equiv \frac{1}{\epsilon} \sum_j \epsilon_j s_j,
\end{equation}
where $s_j$ are the grain sizes for the individual phases. The effective stopping time assuming Epstein drag then corresponds to the usual expression
\begin{equation}
T_{\rm s} \equiv \frac{\rho_{\rm grain} s}{\rho c_{\rm s}},
\end{equation}
where $\rho_{\rm grain}$ is the intrinsic grain density. The upshot of using (\ref{eq:tsj}) is that the implementation of the algorithm is then very similar to \cite{pricelaibe15}, aside from the infrastructure changes required to store and visualise multiple grain populations.

\subsection{Positivity of the dust fraction}
 We do not evolve the dust fraction directly. This is because in \cite{pricelaibe15} we found that errors in the discrete version of (\ref{eq:depsdt}) could produce negative dust densities. This occurs because one requires the symmetric SPH derivative operator in order to conserve the total dust mass. The solution proposed in \cite{pricelaibe15} was to evolve the variable $S \equiv \sqrt{\rho \epsilon}$, from which one can reconstruct the dust fraction via $\epsilon = S^2/\rho$ such that positivity is guaranteed. More recently the issue of maintaining $0 < \epsilon < 1$ was studied in more detail by \cite{ballabioetal18}. They found it was better to use the square root of the dust-to-gas ratio, $\sqrt{\rho_{\rm d}/\rho_{\rm g}}$, as the evolved variable since this guarantees not only positivity but also ensures that $\epsilon < 1$. See \cite{ballabioetal18} for details.
 
 We follow a similar procedure, but evolve the variable $\theta_j \equiv \sin^{-1} (\sqrt{\epsilon})$ such that $\epsilon = \sin^2 \theta$. Written in terms of this variable, (\ref{eq:depsjdt}) becomes
\begin{equation}
	\frac{\mathrm{d} \theta_j}{\mathrm{d} t} = -\frac{1}{2 \rho \sin{\theta_j} \cos{\theta_j}} \nabla \cdot \left( \sin^2{\theta_j} \tilde{t}_{{\rm s}, j} \nabla P \right).
	\label{eq:dthetajdt}
\end{equation}


\subsection{SPH discretisation}
 SPH discretisation of (\ref{eq:drhodtmulti})--(\ref{eq:dudt}) is straightforward since they are almost identical to the usual equations of compressible hydrodynamics. We use
\begin{align}
\rho_a & = \sum_b m_b W_{ab} (h_a); \hspace{1cm} h_a = \eta \left(\frac{m_a}{\rho_a} \right)^{1/\nu}, \label{eq:rhosum} \\
\frac{{\rm d} \bm{v}_a}{{\rm d} t} & = -\sum_b m_b \left[ \frac{P_a + q^a_{ab}}{\Omega_a \rho_a^2} \nabla W_{ab}(h_a) + \frac{P_b + q^b_{ab}}{\Omega_b \rho_b^2} \nabla W_{ab}(h_b) \right], \\
\frac{{\rm d} \tilde{u}_a}{{\rm d} t} & = \sum_b m_b \frac{P_a + q^a_{ab}}{\Omega_a \rho_a^2} (\bm{v}_{a} - \bm{v}_b) \cdot \nabla W_{ab}(h_a),
\end{align}
where $W$ is the usual (bell-shaped) SPH kernel (we employ either the cubic or quintic spline kernel by default), we solve the two equations (\ref{eq:rhosum}) iteratively using a Newton-Raphson solver as described in \cite{pricemonaghan07}, $\Omega$ is the usual correction term related to the gradient of the kernel with respect to the smoothing length (c.f.~\cite{monaghan02}) and we apply the usual artificial viscosity term modified only by the gas fraction, namely
\begin{equation}
q^{a}_{ab} =
\begin{cases}
-\frac{1}{2} \left( 1 - \epsilon_{a} \right) v_{\mathrm{sig},a} \bm{v}_{ab} \cdot \hat{\bm{r}}_{ab}, & \qquad \bm{v}_{ab} \cdot \hat{\bm{r}}_{ab} < 0 \\      
 0,  & \qquad \mathrm{otherwise}.
\end{cases}
\end{equation}
We also include an artificial conductivity term, details of which can be found in \cite{hpl18}.

 We discretise (\ref{eq:dthetajdt}) using
 \begin{align}
	\frac{\mathrm{d} \theta_{j,a}}{\mathrm{d} t}  & =  - \frac{1}{\sin{(2\theta_{j,a})}} \times \nonumber \\ 
	& \sum_b m_b \frac{\sin{\theta_{j,a}} \sin{\theta_{j,b}}}{\rho_a \rho_b} (\tilde{t}_{{\rm s}, j, a} + \tilde{t}_{{\rm s}, j, b}) (P_a - P_b) \frac{\overline{F}_{ab}}{|r_{ab}|},
	        \label{eq:sph_dustfrac}
\end{align}
%
where $\overline{F}_{ab} \equiv \frac{1}{2}[ F_{ab}(h_{a}) + F_{ab}(h_{b}) ]$ and $F_{ab}$ is the scalar part of the kernel gradient, such that $\nabla_{a} W_{ab} \equiv \hat{\bm{r}}_{ab} F_{ab}$.

\subsection{Timestep constraint}
A linear perturbation analysis shows that the timestep stability condition for a mixture described by (\ref{eq:drhodtmulti})--(\ref{eq:depsjdt}) is given by
\begin{equation}
\Delta t < C_0 \frac{h}{\sqrt{\tilde{c}^2_{\rm s} + \epsilon^2 T_{\rm s}^2 c_{\rm s}^4/ h^2}}.
\end{equation}
The first term in the denominator is the usual Courant condition involving the modified sound speed. The second term represents a parabolic timestep constraint, i.e. $\Delta t < C_0 h^2 / \kappa$, where $\kappa \equiv \epsilon^2 T_{\rm s}^2 c_{\rm s}^2$. A parabolic constraint may seem prohibitive, but only becomes important when the effective stopping time is \emph{long}, which is when the terminal velocity approximation (which we have assumed) is no longer valid. In this regime the usual method of representing the gas and dust as separate sets of particles is more appropriate.

\begin{figure*}
\begin{center}
\includegraphics[width=\textwidth]{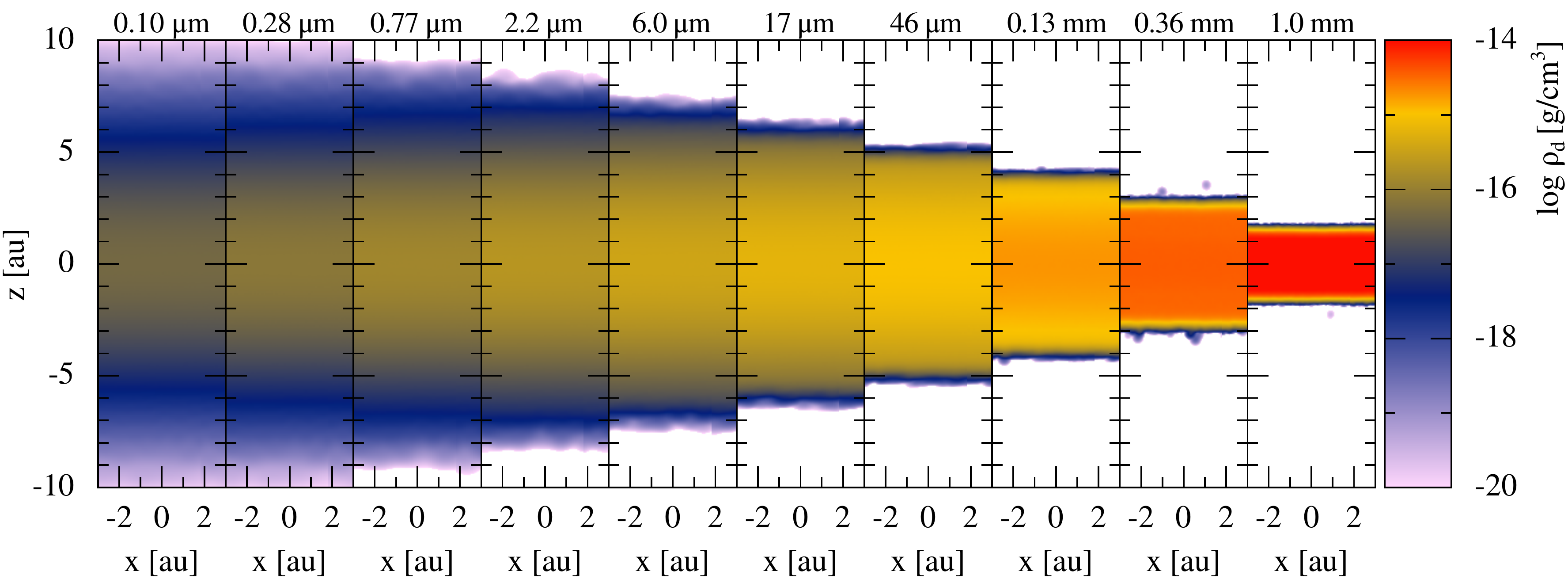}
\caption{Ten dust densities from a MULTIGRAIN simulation after having settled for 15 orbits in a 3D vertical column of a protoplanetary disc using $100 \times 86 \times 78 = 670,800$ simulation particles. Our MULTIGRAIN simulation is $\sim 5\times$ faster to run than 10 single-phase simulations run serially. Reproduced from \cite{hpl18}.}
\label{fig:settling}
\end{center}
\end{figure*}
  
\begin{figure}
\begin{center}
\includegraphics[width=\columnwidth]{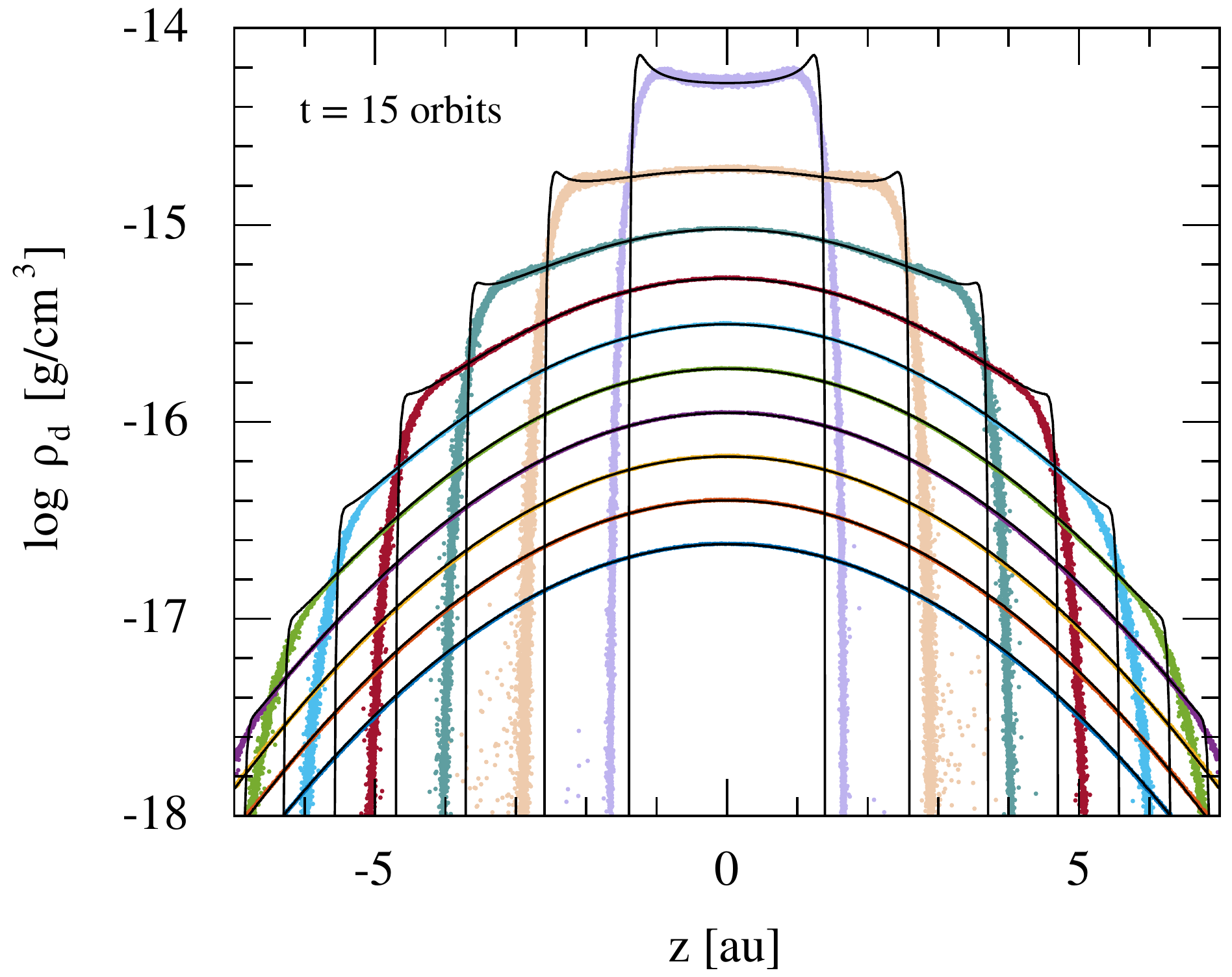}
\caption{Comparison of the dust densities for the different grain populations in our SPH dust settling test to reference solutions computed by solving the 1D differential equations for each grain population (solid lines). Reproduced from \cite{hpl18}.}
\label{fig:settlinglines}
\end{center}
\end{figure}

\section{Numerical tests}
 One of the problems when dealing with gas-dust mixtures at all is that there are few analytic solutions available in the literature with which to benchmark a numerical code. This led us to publish our own analytic solutions for wave \cite{laibeprice11} and shock \cite{laibeprice12} problems. With multiple dust populations the situation is even worse, although there have been some attempts to derive steady state solutions for multiple grain populations in protoplanetary discs \cite{baistone10a}.
 
\subsection{Sanity check: Do we recover analytic solutions for single phase dust-gas mixtures if we split the grain population?}
Our first series of tests utilised the analytic solutions for the wave, shock and dust diffusion problems employed by \cite{pricelaibe15}. Since these are only applicable to single phase mixtures, we performed these tests mainly as a sanity check. That is, we split our previous single population of grains into 10 sub-populations each with one tenth of the previous (total) dust density. Each grain population therefore contains one tenth of the original grains but with each sub-population having the same grain size. Figure~\ref{fig:dusttest} shows the results of this test, confirming that our numerical solutions with the {\sc multigrain} algorithm correctly reduce to the corresponding single grain solutions, and that these in turn match the analytic solutions (solid lines).

\subsection{Settling of multiple grain populations in a protoplanetary disc}
 Our second series of tests was performed using a simplified version of settling in a protoplanetary disc. For this problem we consider only the vertical component of the gravitational force from the central star. We then set up a hydrostatic atmosphere of gas plus several populations of grains with density profiles matching the initial gas density profile, namely
\begin{equation}
\rho_{\rm g}(z) = \rho_{{\rm g}, 0} \exp \left[-\frac{z^2}{2 H^2} \right],
\end{equation}
where $H = c_{\rm s}/\sqrt{GM/R^3}$ is the pressure scale height in the $z$ direction. We perform the test in a 3D cartesian geometry, with periodic boundary conditions in the $x$ and $y$ directions and free boundaries in the $z$ direction. For this calculation we again employed 10 different grain populations with sizes ranging from 0.1$\mu$m to 1 mm. We adopt physical parameters appropriate to conditions in protoplanetary discs, with $M$ equal to the mass of the Sun, $R = 50$ times the Earth-Sun distance and $\rho_{{\rm g}, 0} \approx 6 \times 10^{-13}$ g/cm$^3$. We set the sound speed such that $H/R = 0.05$. We employ an Epstein drag prescription for the grains. 

 Figure~\ref{fig:settling} shows the resulting grain evolution. The smallest grains (left panels) remain glued to the gas, while the larger grains progressively decouple from the gas and settle to the disc midplane. Importantly, the SPH particles are almost at rest in this problem --- the dust evolution is governed almost entirely by the changing dust fractions. While the differences are qualitative, this reflects our expectation regarding the behaviour of the different grain populations. The results are similar to --- but not identical --- to the results when each population is simulated individually. The results are not identical because of the different backreaction induced on the gas when there are multiple grains compared to only a single grain species.
 
  Figure~\ref{fig:settlinglines} shows a more quantitative evaluation of this test, comparing the evolution of the dust fractions with our SPH algorithm to reference solutions computed by solving the 1D differential equation for the settling of each grain size in the vertical direction. The locations of the settling fronts from our {\sc multigrain} code match the reference solutions to better than a few percent for all except the largest grain sizes.
   
  Further tests of the method, including a comparison to the analytic solution derived by \cite{baistone10a} can be found in \cite{hpl18}. The remarkable aspect of our new {\sc multigrain} method is that we are now able to capture in a single simulation what would have previously required 10 or more individual simulations. Furthermore, those 10 simulations would miss potentially important effects from the global backreaction of the grains onto the gas.

\begin{figure}
\begin{center}
\includegraphics[width=\columnwidth]{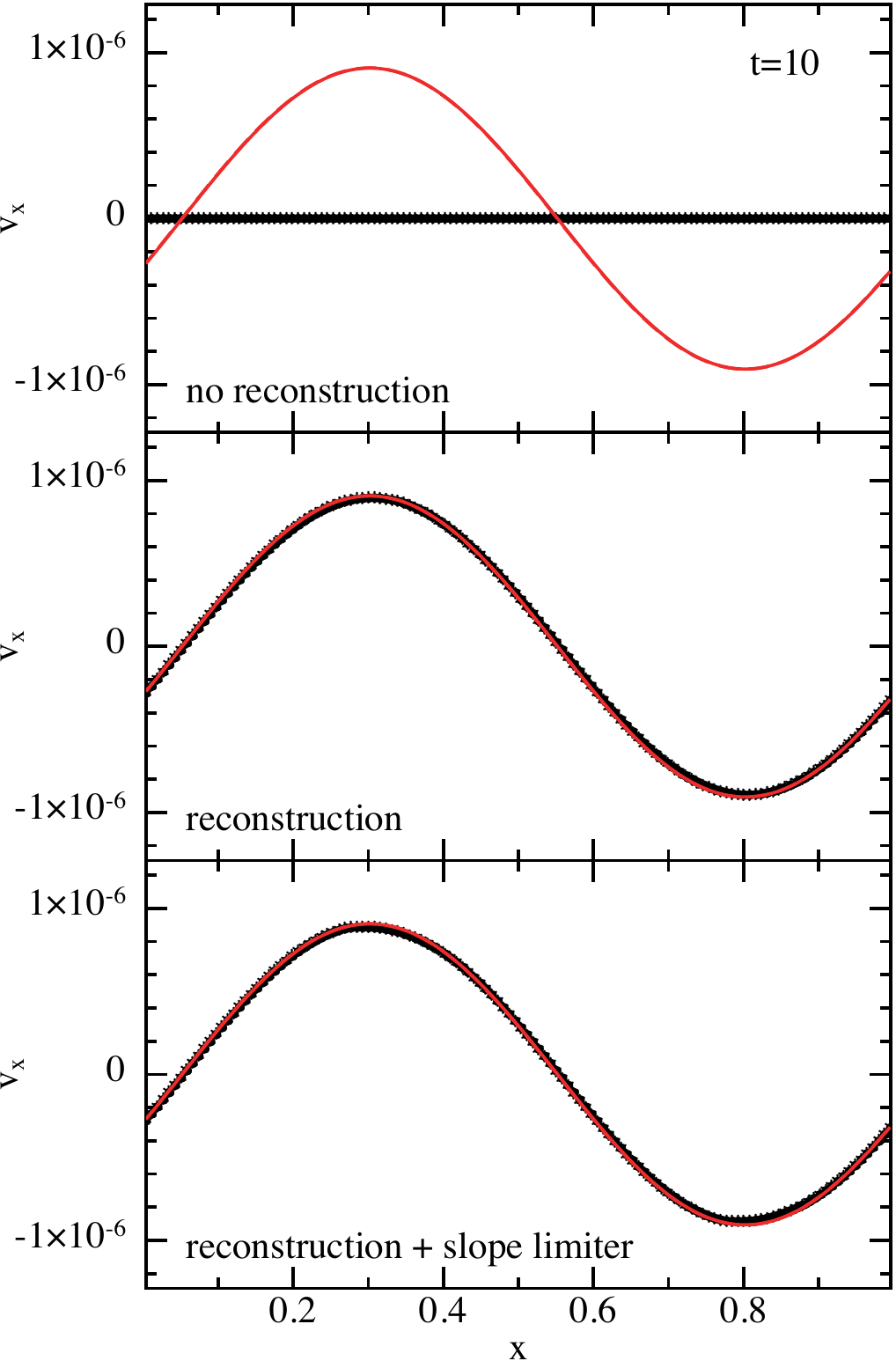}
\caption{Solution to the {\sc dustywave} problem with strong drag ($K=1000$) when using separate sets of gas and dust SPH particles. The SPH solution is shown by the black dots, while the analytic solution is given by the red line. According to (\ref{eq:disp}) there should be little damping of the wave, but applying (\ref{eq:vgsph})--(\ref{eq:vdsph}) produces an overdamped solution. Using reconstructed velocities (bottom two panels) solves this problem and avoids the need for prohibitive spatial resolution.}
\label{fig:dustywave}
\end{center}
\end{figure}

\begin{figure}
\begin{center}
\includegraphics[width=\columnwidth]{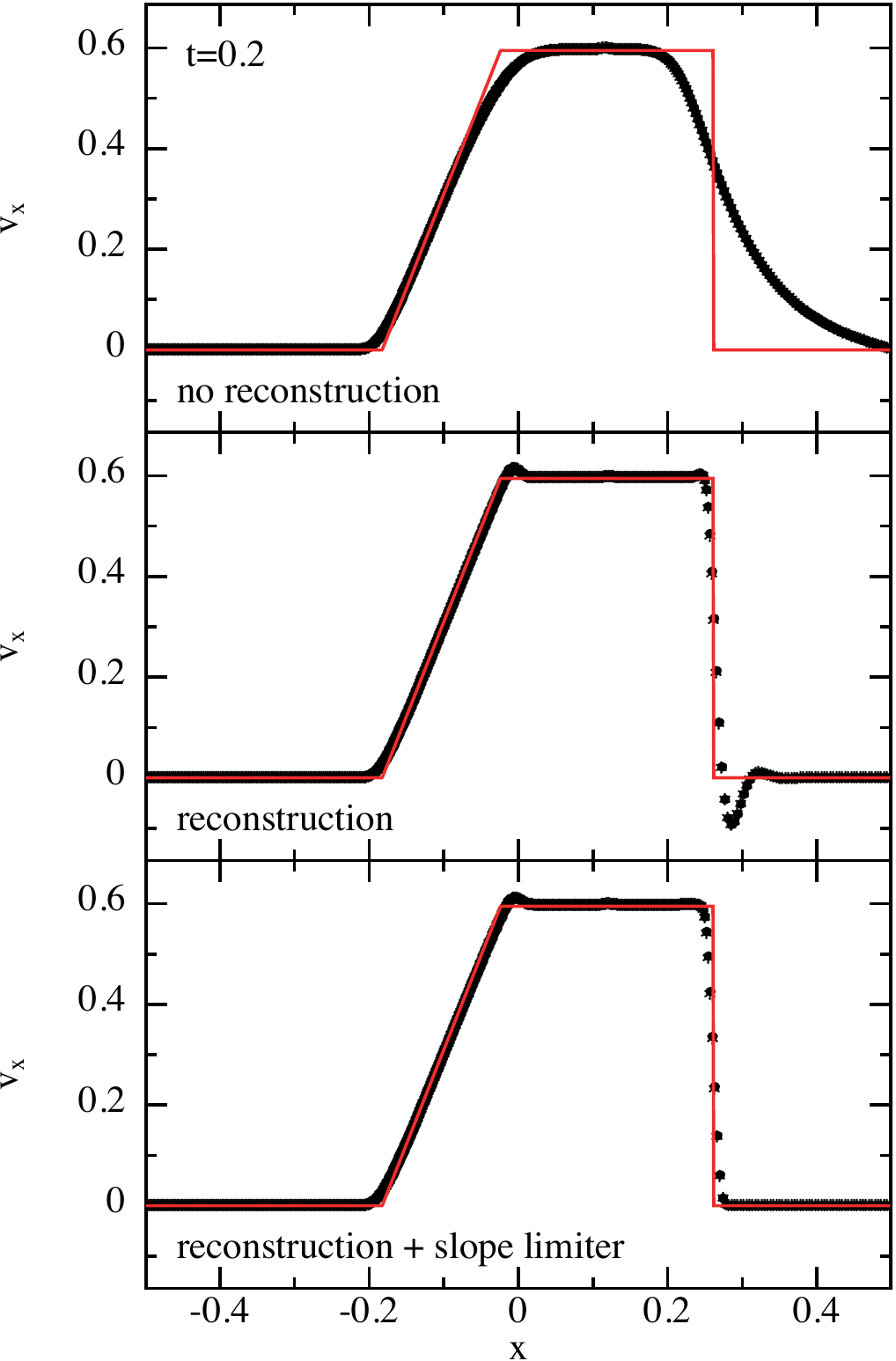}
\caption{As in Figure~\ref{fig:dustywave} but for the {\sc dustyshock} problem. Again, the SPH solution computed with separate gas and dust particles (top; black points) is highly inaccurate compared to the analytic solution (red line). Applying our reconstruction algorithm solves the overdamping problem but also introduces spurious oscillations (middle panel). These can be eliminated by employing a slope limiter (bottom panel).}
\label{fig:dustyshock}
\end{center}
\end{figure}

\section{Solving the overdamping problem}
\label{sec:overdamping}
 The last piece of the puzzle is to increase the degree of crossover between our `gas and dust with separate sets of SPH particles' and `multiple-dust-grains-and-gas mixture' methods, so that it is possible to obtain accurate results for any value of the stopping time with either method. For large grains, one of the major issues is the overdamping problem discussed in Section~\ref{sec:problems}. 
 
\subsection{Interpolating gas and dust velocities}
Recently, we found that the overdamping problem can be solved by replacing the velocities of the gas and dust particles used in the drag terms (\ref{eq:vgsph})--(\ref{eq:vdsph}) with velocities `reconstructed' to the barycentre of each pair of particles. That is, we use
\begin{align}
\bm{v}^{*}_a & = \bm{v}_a + \left(\bm{r}^{*} - \bm{r}_{a}\right)^\beta \frac{\partial \bm{v}_{a}}{\partial \bm{r}_{a}^\beta}.\\
\bm{v}^{*}_i & = \bm{v}_i + \left( \bm{r}^{*} - \bm{r}_{i}\right)^\beta \frac{\partial \bm{v}_{i}}{\partial \bm{r}_{i}^\beta}.
\end{align}

At the barycentre $\bm{r}^{*} = \bm{r}_{a} + 0.5 \bm{r}_{ai}$, these relations combine to
\begin{equation}
\bm{v}^{*}_{ai} \cdot \hat{\bm{r}}_{ai} = \bm{v}_{ai} \cdot \hat{\bm{r}}_{ai} + 0.5 \vert r_{ai} \vert  \left( S_{ai} + S_{ia} \right) ,
\end{equation}
where $S_{ai} \equiv \hat{r}_{ai}^\alpha \hat{r}_{ai}^\beta \frac{\partial v_{a}^{\alpha}}{\partial x_{a}^\beta}$. We then replace $\bm{v}_{ai} \cdot \hat{\bm{r}}_{ai}$ with $\bm{v}^{*}_{ai} \cdot \hat{\bm{r}}_{ai}$ in equations (\ref{eq:vgsph})--(\ref{eq:dusph}). Velocity gradients required for the reconstruction are computed during the density summation according to
\begin{equation}
\frac{\partial v_a^{\alpha}}{\partial r_a^{\beta}}  = - \frac{1}{\Omega_{a}\rho_{a}} \sum_{b}m_{b} v^{\beta}_{ab}\nabla^{\beta}W_{ab}\left(h_{a} \right).
\end{equation}
Finally, to avoid introducing new maxima or minima, we limit the factor $0.5\left(S_{ai} + S_{ia}\right)$ using a slope limiter, i.e. a function $f\left(S_{ai},S_{ia} \right)$ that prevents the development of numerical oscillations. We use the Van Leer MC limiter, since it provides best compromise between smoothing and dissipation.

\subsection{Numerical tests}
Figure~\ref{fig:dustywave} shows the results of the {\sc dustywave} test from \cite{laibeprice11} when separate sets of gas and dust particles are employed and the drag interaction is computed via (\ref{eq:vgsph})--(\ref{eq:vdsph}). The problem is set up in 1D with $\rho_{{\rm g}, 0} = \rho_{{\rm d}, 0} = 1$, $c_{{\rm s}, 0} = 1$ a constant drag coefficient $K = 1000$ and sinusoidal perturbations to the density, velocity and thermal energy with amplitude $10^{-6}$. The analytic solution is given by the red lines, while the SPH solution is shown by the black points. The top panel illustrates the `overdamping problem' --- the amplitude of the wave is severely damped compared to the analytic solution. Applying our new reconstruction algorithm resolves this issue (middle and lower panels) and avoids the need for the prohibitive spatial resolution criterion (\ref{eq:hcsts}).

 Figure~\ref{fig:dustyshock} is similar but showing the results of the {\sc dustyshock} test from \cite{laibeprice12} with strong drag ($K=1000$). Again, applying reconstruction solves the overdamping problem but in the absence of a slope limiter also introduces spurious oscillations into the solution (middle panel). Using the slope limiter eliminates these (lower panel) giving a solution to within a few percent of the analytic solution (red line).

\section{Conclusion}
 In summary, we have extended our SPH algorithms for dust-gas mixtures to the case of multiple small grain populations mixed with gas. This will enable us to study the evolution of interstellar dust in protoplanetary discs and in the dusty clouds from which stars are born. Our {\sc multigrain} algorithm is implemented and available to download in our publicly available SPH code, \textsc{Phantom} \cite{priceetal18a}.
 
 Much work remains. In particular we are working towards an algorithm to self-consistently model grain growth and destruction within our {\sc multigrain} code, and towards a hybrid code capable of simulating the full range of grain sizes in protoplanetary discs, from micron-sized dust to km-sized boulders, self-consistently. These will have to wait for a future SPHERIC!

\section*{Acknowledgments}
 DP and MH acknowledge funding from the Australian Research Council via DP130802078 and FT130100034.

\bibliographystyle{IEEEtran.bst}
\bibliography{dan}

\end{document}